\documentclass[twocolumn,prl,aps,floats,floatfix,showpacs,
               superscriptaddress]{revtex4}
\usepackage{graphicx}
\usepackage{bm}

\begin{document}
%
%
\newcommand{\be}{\begin{equation}}
\newcommand{\ee}{\end{equation}}
\newcommand{\bea}{\begin{eqnarray}}
\newcommand{\eea}{\end{eqnarray}}
\newcommand{\beann}{\begin{eqnarray*}}
\newcommand{\eeann}{\end{eqnarray*}}
\newcommand{\bma}{\begin{array}{cc}}
\newcommand{\ema}{\end{array}}
\newcommand{\fr}{\frac}
\newcommand{\ra}{\rangle}
\newcommand{\la}{\langle}
\newcommand{\li}{\left}
\newcommand{\re}{\right}
\newcommand{\ri}{\right}
\newcommand{\uarr}{\uparrow}
\newcommand{\darr}{\downarrow}
\newcommand{\df}{\stackrel{\rm def}{=}}
\newcommand{\nn}{\nonumber}
\newcommand{\dpl}{\displaystyle}

\newcommand{\alp}{\alpha}
\newcommand{\sig}{\sigma}
\newcommand{\eps}{\epsilon}
\newcommand{\xsi}{\xi}
\newcommand{\lam}{\lambda}
\newcommand{\ny}{\nu}

\newcommand{\HamO}{\hat{H}_0}
\newcommand{\Ham}{\hat{H}}
\newcommand{\HamV}{\hat{V}_c}
\newcommand{\seteps}{ \{ \eps \} }
\newcommand{\setlam}{ \{ \lam \} }
\newcommand{\ef}{E_F}
\newcommand{\Deltaml}{d}
\newcommand{\Deltaov}{\Delta}
\newcommand{\Deltamubargamma}{ \Deltaov_{\bar{\mu}\gamma} }
\newcommand{\Deltaibarj}{ \Deltaov_{\bar{i} j} }
\newcommand{\vc}{v_c}
\newcommand{\VKOSTYA}{V_K}
\newcommand{\delEF}{\delta_F}

\newcommand{\br}{{\bf r}}

%
%
\title{Fermi-Edge Singularities in the Mesoscopic X-Ray Edge Problem}

\author{Martina Hentschel}
\affiliation{
Department of Physics, Duke University, Box 90305, Durham, NC 27708-0305}

\author{Denis Ullmo}

\affiliation{
Department of Physics, Duke University, Box 90305, Durham, NC 27708-0305}
\affiliation{
Laboratoire de Physique Th\'eorique et
Mod\`eles Statistiques (LPTMS), 91405 Orsay Cedex, France}

\author{Harold U. Baranger}
\affiliation{
Department of Physics, Duke University, Box 90305, Durham, NC 27708-0305}

\date{\today}

%
%
\begin{abstract}
{ We study the x-ray edge problem for a chaotic quantum dot or nanoparticle
displaying mesoscopic fluctuations. In the bulk, x-ray physics is known to
produce Fermi edge singularities -- deviations from the naively expected
photoabsorption cross section in
the form of a peaked or rounded edge.  For a coherent system with chaotic
dynamics, we find substantial changes; in particular, a photoabsorption
cross section showing a rounded edge in the bulk will change to a slightly
peaked edge on average as the system size is reduced to a mesoscopic (coherent)
scale.}
\end{abstract}
\maketitle


The x-ray edge problem for metals refers to a singularity at threshold in the
photoabsorption spectra associated with the excitation of a core electron to the
conduction band by an x ray.  The singularity takes the form of a peaked or
rounded edege and was studied intensively over the last four decades using
different methods~\cite{tanabe:RMP1990}. The sustained attention devoted to this
problem developed because the basic phyiscs originally studied for x-ray
absorption recurs in many other contexts. Recently, photoabsorption in
semiconductor quantum wells \cite{qwells} and electrical conduction through
localized states \cite{localizedIV} have been particularly active areas of Fermi
edge research. Motivated by the growing interest in nanoscale systems like
quantum dots or nanoparticles
\cite{sohn:MesoBook1997,alhassid:RMP2000,RalphvDelft01}, we address Fermi edge
physics for such confined geometries, assuming generic chaotic dynamics of the
electrons.

For bulk metals or degenerately doped semiconductors, the photoabsorption
cross section $A(\omega)$ near
the threshold energy $\omega_{\rm th}$ involves two competing effects,
Mahan's enhancement and Anderson's orthogonality catastrophe
(AOC). Its form is \cite{tanabe:RMP1990}
\begin{equation}
A(\omega) \propto (\omega-\omega_{\rm th})^{-2 \li( |\delta_{l_0}|/\pi \re)
+ \sum_{l} 2 (2 l+1) \li( \delta_l / \pi \re)^2} \:.
\label{Aofw_thres}
\end{equation}
Here $\delta_l$ is the partial-wave phase shift at the Fermi energy
associated with the core hole potential $\HamV$ for orbital channel
$l$; $l_0$ is the optically excited channel.  The phase shifts obey
the Friedel sum rule $Z=\sum_l 2 (2 l+1) \delta_l/\pi$ with $Z=-1$
here.  AOC \cite{anderson:PRL1967} reduces the overlap between ground
states before and after $\HamV$ is applied, leading to a rounded edge.
On the other hand, the sudden presence of a charge associated with the
missing core electron entails a many body response that enhances the
threshold cross section, resulting in a peaked edge.  Whereas all $l$
channels are sensitive to $\HamV$ and so contribute to AOC (second
term in the exponent), only the optically excited $l_0$ experiences
the many-body interaction (first term).

The many-body enhancement depends, via the dipole selection rule, on
the symmetry of both the core electron and local conduction
electron wavefunctions.  We will assume the latter to be of $s$-type,
and distinguish between core electrons with $s$-symmetry (K-shell) and
$p$-symmetry (L$_{2,3}$-shell) \cite{citrin:PRB1979}, referring to the
photoabsorption threshold as a K- or L-edge, respectively.
Furthermore, we take the core hole potential to be spherically
symmetric: $\delta_l = 0$ for $l \neq 0$.  The Friedel sum rule then
implies $ \delta_0 = - \pi/2$, so that either a rounded K-edge ($l_0=1$,
photoabsorption dominated by AOC) or a peaked L-edge ($l_0=0$) is
found.  For mesoscopic systems, in addition to $\delta_0 \approx
-\pi/2$, we shall consider smaller values for situations where other
charges (e.g. on nearby gates in a semiconductor heterostructure) help
to screen the core hole.

Compared to bulk metals, some characteristics intrinsic to nanoscale objects
will significantly alter the x-ray edge properties.  First, the fact that both
the number of electrons and the total number of levels are finite will lead to
an incomplete AOC and many-body enhancement.  Furthermore, the lack of
rotational symmetry eliminates independent orbital momentum
channels. Finally, mesoscopic fluctuations will affect both energy levels and
wavefunctions. In tackling these issues, the key ingredient is the joint
statistics of the electronic properties before and after the core hole
excitation \cite{matveev:PRL1998}.

The influence of these effects on the average and fluctuations of the
Anderson overlap was modeled using parametric random matrices
\cite{vallejos:PRB2002} and for disordered systems
\cite{gefen:PRBR2002}.
These studies mainly addressed the regime where the added potential
can be treated perturbatively, as, for instance, for
scrambling phenomena in Coulomb blockade experiments. Here, we study first
AOC and then absorption spectra near threshold for a model
corresponding to the x-ray edge problem.
In this case the perturbing potential $\HamV$ associated with the core hole has
a very short range, but on the other hand strong perturbations are physically relevant.
Possible experimental realizations are briefly discussed in the conclusion.

We model the conduction electrons
of our mesoscopic system in the absence of the core hole by
the non-interacting $ \HamO = \sum_{i , \sigma} \eps_i c^\dagger_{i,\sigma}
c_{i,\sigma}$, where $c^\dagger_{i,\sigma}$ creates a particle with spin $\sigma
= \pm $ in the orbital $\varphi_i(\br)$ ($i=0,\ldots,N-1$).  We furthermore
assume that the perturbing potential
is a contact potential $\HamV = {\cal V} \vc
|\br_c \ra \la \br_c |$, with $\br_c$ the location of the core hole and ${\cal V}$
the volume of the system. The diagonal form of the
perturbed Hamiltonian is $\Ham = \HamO + \HamV = \sum_{i , \sigma} \lambda_i
\tilde c^\dagger_{i,\sigma} \tilde c_{i,\sigma}$, where $\tilde
c^\dagger_{i,\sigma}$ creates a particle in the perturbed orbital $\psi_i(\br)$.
For the Fermi energy in the middle of the conduction band, $\delta_0$, $\vc$ and
the mean level spacing $\Deltaml$ are related through
$\delta_0= \arctan (\pi {\vc}/{\Deltaml})$ \cite{tanabe:RMP1990}
($\delta_0$ is negative since the core potential is attractive).

A  remarkable  property of a rank  one  perturbation such as a contact potential
is that  all  the quantities of interest  for the  x-ray  edge problem  can be
either expressed  in  terms of  the $\seteps$  and $\setlam$ or   taken as
independent random variables.  For instance, ignoring for now the spin variable,
the  overlap between the  many body ground states  with $M$ particles $\Phi_0$
and $\Psi_0$ of $\HamO$ and $\Ham$ is \cite{tanabe:RMP1990}
\begin{equation}
\label{eq:overlap}
|\Deltaov|^2=
|\la \Psi_0 | \Phi_0 \ra |^2 = \prod_{i=0}^{M-1}  \prod_{j=M}^{N-1}
\fr{ (\lam_j - \eps_i)  (\eps_j - \lam_i) }{ (\lam_j - \lam_i)  (\eps_j - \eps_i)} \:.
\end{equation}

For chaotic mesoscopic systems, we may assume that the unperturbed energy levels
follow random matrix theory fluctuations and the wave function intensities are
Porter-Thomas distributed (in particular at $\br_c$)
\cite{bohigas:LesHouches91}.  Under these hypotheses, the joint probability
distribution $P(\seteps,\setlam)$ was derived by Aleiner and Matveev
\cite{matveev:PRL1998} (see \cite{smolyarenko:PRL2002} for a generalization to
the non-rank-one case):
\begin{equation}
  P(\seteps,\setlam) \propto
  \frac{\prod_{i>j} (\eps_i - \eps_j) (\lambda_i - \lambda_j)}
       {\prod_{i,j} \left|\eps_i - \lambda_j \right|^{1-\beta/2}}
       e^{ - \frac{\beta}{2} \sum_i
             (\lambda_i - \eps_i)/v_i }
  \label{eq:kostya}
\end{equation}
with the constraint $\eps_{i-1} \le \lambda_i \le \eps_i$.  Here $\beta = 1$
(circular orthogonal ensemble, COE) or $\beta=2$ (circular unitary ensemble,
CUE) for the time reversal symmetric or asymmetric case, respectively. In the
middle of the band $v_i =\vc$.  In our case, it turns out to be necessary to
include boundary effects \cite{inprep}, which modify the phase shifts
away from the band center. This can be done simply by using a variable $v_{i}$
given by
\begin{equation}
\fr{1}{v_i}=\fr{1}{\vc} + \fr{1}{\Deltaml} \ln \fr{N-0.5-i}{i+0.5}
                        \:.
                        \nonumber
\end{equation}
for $i \in [0,(N-1)/2]$, and the analogous form for $i \in [(N-1)/2,N-1]$. In addition, at
perturbation strengths $\li|\vc \re| \gtrsim 0.5 \Deltaml$, the lowest perturbed
level $\lam_0$ requires special treatment as its mean shift $\langle
\lam_0 - \eps_0 \rangle = - N \Deltaml / [\exp(\li| \Deltaml/\vc \re|) - 1]$
becomes large. This gives rise to a second band in the x-ray spectra
\cite{tanabe:RMP1990} that we do not discuss here.

Using a Metropolis algorithm, we can therefore generate ensembles of
$(\seteps,\setlam)$ with the proper joint distribution, Eq.~(\ref{eq:kostya}),
and using Eq.~(\ref{eq:overlap}) build the distribution of overlap for any value
of the parameters $M$, $N$, $\vc$, and $\beta$.  To quantify the role of
fluctuations, we define for reasons of comparison the special case of
equidistant unperturbed levels $\{ \eps \}$, and will
refer to it as the uniform or ``bulklike'' case with overlap $|\Deltaov_b|^2$
($ \propto N^{- \delta_F^2/\pi^2}$).

\begin{figure}[b]
\includegraphics[width=8.5cm]{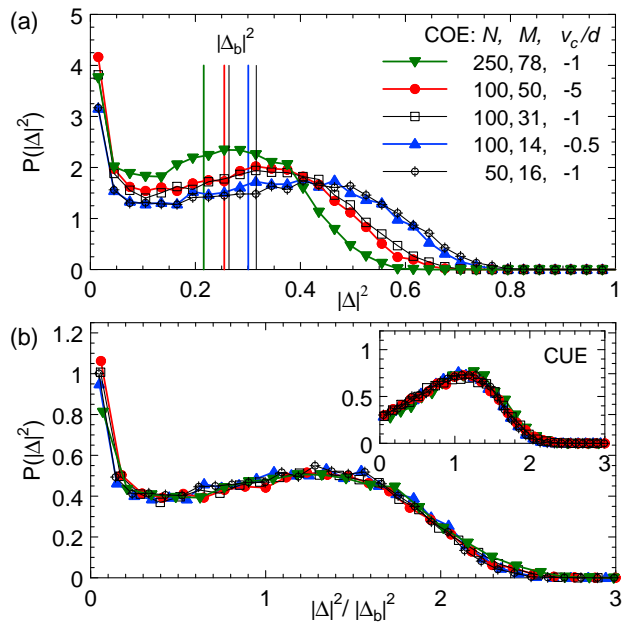}
\caption{(Color online) Probability distribution $P(|\Deltaov|^2)$
of the overlap $|\Deltaov|^2$ for
different parameter sets $\{N,M,-v_c/\Deltaml\}$.
  (a) Although all $\delEF \approx -\pi/2$,
  the $P(|\Deltaov|^2)$
  are visibly different as a result of different bulklike
  overlaps $|\Deltaov_b|^2$ (increasing
  with the legend entries, COE case).
  (b) The curves coincide
  when scaled with $|\Deltaov_b|^2$. The inset shows the CUE result with a
  noticeably different $P(|\Deltaov|^2)$ distribution
  for small $|\Deltaov|^2$ originating from the
  different wavefunction statistics.
  \label{fig_same_delf_comb}
  }
\end{figure}

\begin{figure}
\includegraphics[width=8.5cm]{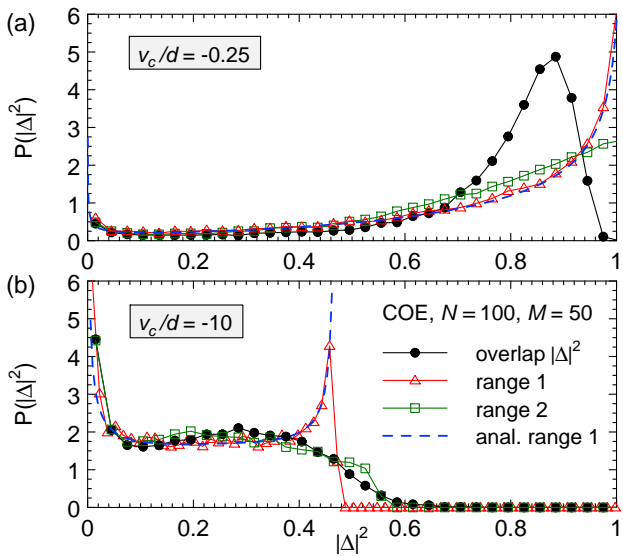}
  \caption{(Color online)
  Exact and approximate probability distributions of the overlap
  for (a) weak ($v_c/\Deltaml=-0.25$)
  and (b) strong ($v_c/\Deltaml=-10$) perturbation (COE, $N=100, M=50$).
  The exact $P(|\Deltaov|^2)$ is increasingly well reproduced
  by the range-1 and range-2 approximation,
  indicating that the
  fluctuations originate from levels around the Fermi energy.
  The dashed line denotes the range-1 analytical result.
  \label{fig_overl_order123}
  }
\end{figure}

Results of such a simulation are displayed in Figs.~\ref{fig_same_delf_comb}
and \ref{fig_overl_order123}. The probability distributions of the overlap
$P(|\Deltaov|^2)$ shown in Fig.~\ref{fig_same_delf_comb}(a) differ considerably
even though the phase shift at the Fermi energy, $\delta_F \equiv \arctan ( \pi
{v_{i=M}}/{\Deltaml})$, is similar. In fact, as expected, we find that both
$P(|\Deltaov|^2)$ and the bulklike result $|\Deltaov_b|^2$ depend independently
on $N$, $M$, and $\delta_F$.  However, performing a scaling by $|\Deltaov_b|^2$,
see Fig.~\ref{fig_same_delf_comb}(b), causes all curves with the same $\delta_F$
to collapse. \textit{Thus, the fluctuations of the overlap depend only on the
value of the phase shift at the Fermi energy, $\delta_F$.}

The small $|\Deltaov|^2$ part of $P(|\Deltaov|^2)$ can be accurately
reproduced, as in the cases studied in
Refs.~\cite{gefen:PRBR2002,vallejos:PRB2002}, by incorporating
the fluctuations of only the levels closest to the Fermi energy. For
instance, we may multiply $|\Deltaov_b|^2$ by the fluctuating factor
$[ 1 /(1 - \delEF /\pi)](\lam_{M+1} - \eps_M)/(\lam_{M+1}-\lam_M) $,
which we call the ``range-1'' approximation.  The resulting
distribution (Fig.~\ref{fig_overl_order123}) is already similar to the exact one.
We found an analytical result for
this approximation for both the COE and CUE cases by treating
two fluctuating levels and reducing  the remaining ones to a single
Gaussian variable \cite{inprep}; it is seen to
coincide with the range-1 result over the whole range of
$|\Deltaov|^2$.

We can similarly define ``range-$n$'' approximations which include the
fluctuations of $n$ levels on each side of the Fermi energy.  The
range-2 approximation matches the exact one for large perturbation
strengths (Fig.~\ref{fig_overl_order123}), but does not improve
significantly on the range-1 approximation for weak
perturbations. We have checked, however, that even in this case,
it is sufficient to treat a few levels around $E_F$ exactly to obtain
the correct $P(|\Deltaov|^2)$.  This retrospectively justifies our use
of a full random matrix to model our
chaotic systems -- the levels beyond the Thouless energy, where such a
model does not apply \cite{ABG02}, do not affect the overlap
distribution.

Turning now to the absorption spectra, we limit our discussion to  the K-edge
($s$-symmetry of the core electron wavefunction) and approach the mesoscopic
x-ray edge problem using Fermi's golden rule \cite{tanabe:RMP1990},
\begin{equation}
  A(\omega) =2 \pi \hbar^{-1}
  \sum_f | \la \Psi_f | \hat D | \Phi_0^c \ra |^2
  \delta(E_f - E_0^c - \omega) \:,
\label{Aofw_Fermigoru}
\end{equation}
where the sum is taken over all perturbed final states $\Psi_f$ connected  to
the unperturbed groundstate $\Phi_0^c = \prod_{\sigma=\pm } \prod_{j=0}^{M-1}
c^\dagger_{j,\sigma} |c\ra$ by the dipole operator $\hat D$. ($|c\ra =
c^\dagger_c |0\ra$ is the empty band plus a core electron of spin $\sigma=+$.)
We are interested in processes involving the core hole; thus, the
dipole operator can be written as $\hat D =$ $\mbox{const} \sum_{j=0}^N \li(
w_{j} \tilde{c}_{j,+}^\dagger {c}_c + h.c.  \re)$. Since we assume the core
electron wavefunction and the local part of the conduction electron wavefunction
are both $s$ wave, $w_j$ is related to the derivative of the perturbed orbital
$\psi_j$ in the direction $\vec e$ of the polarization of the x ray through $w_j
= \vec{e} \cdot {\bf \nabla} \psi_j (\br_c)$.

First consider conduction electrons with the same spin as the excited core
electron, and assume $\omega= \omega_{\rm th}$ so that the only possible final
state is $\Psi^+_{f^0} =  \prod_{j=0}^{M}
\tilde{c}_{j,+}^\dagger$.
Without a perturbing potential, the only contribution is the direct process $
w_{M}  \tilde{c}_{M,+}^\dagger {c}_0 $. In the presence of a
perturbation, however, the new and old orbitals are not identical, and terms
with $j < M$ (called replacement processes) also contribute coherently, giving
\cite{tanabe:RMP1990}
\begin{equation}
 |\la \Psi_{f^0} | \hat D | \Phi_0^c \ra |^2  \propto
 |w_{M} \Deltaov|^2   \li| 1 - \sum_{i=0}^{M-1}
                           \fr{ w_{i } \Deltaov_{\bar i, M}}
                              { w_{M} \Deltaov }  \re | ^2
\label{dirrepl}
\end{equation}
where $\Deltaov_{\bar i, M}$ is defined by generalizing Eq.~(\ref{eq:overlap})
with level $i \, (<M)$ replaced by $M$.  Since for chaotic systems the
derivative of the wavefunction, $ {\cal V} k^{-2} \times | \nabla_{\vec{e}}
\psi_j|^2$, is known to have Porter-Thomas fluctuations uncorrelated with the
wavefunction itself \cite{prigodin}, we can proceed as for the overlap to
construct the distribution of $|\la \Psi_{f^0} | \hat D | \Phi_0^c \ra |^2 $.

Away from threshold, part of the x-ray energy can excite additional electrons
above the Fermi energy in so-called shake-up processes.  Their contribution is a
straightforward generalization of Eq.~(\ref{dirrepl}).  The number of these
processes grows in principle exponentially with the energy of the x ray.
However, it remains finite even for large $\omega-\omega_{\rm th}$ (and
not very large for the values of $M$ and $N$ here) if one considers only
shake-up processes that contribute significantly to absorption.
Electrons with spin opposite to the excited core electron
(spectator spin) affect the absorption only through AOC.

\begin{figure}
\includegraphics[width=8.5cm]{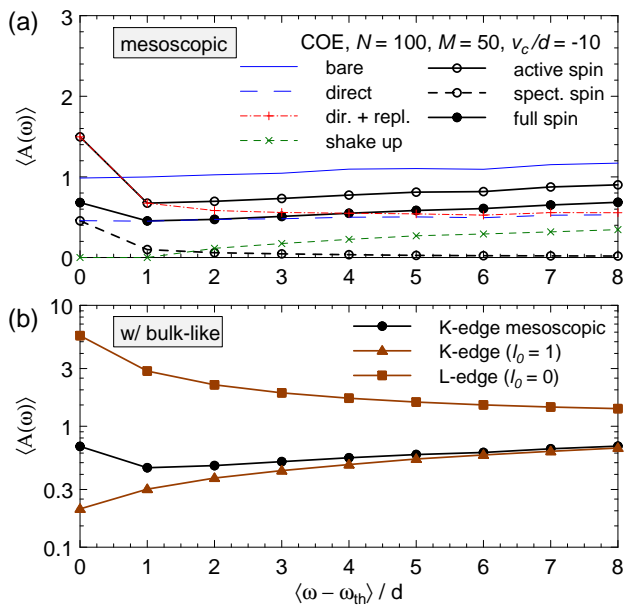}
  \caption{(Color online) X-ray photoabsorption spectra for
  (a) the mesoscopic (chaotic) and
  (b) the bulklike (non-fluctuating) case in terms of averaged
  cross sections (COE, $N=100, M=50, \vc/\Deltaml =-10$).
  (a) The mesoscopic spectrum (filled circles) shows a (slightly) peaked edge.
  Open circles indicate the individual spin contributions for
  the optically active (full line)
  and spectator (dashed line) spins.
  The former arises as the sum of direct and replacement (dashed-dotted),
  as well as shake-up (dashed, crosses), processes.
  For comparison we give the bare spectrum (thin line) without either
  AOC or many-body effects, and
  the result with AOC only (direct process, thin dashed line).
  (b) Comparison of the peaked mesoscopic spectrum (circles)
  to the bulklike results for the same optical channel
  (triangles, $l_0=1$) that shows a rounded K-edge. The mesoscopic peak is, however,
  less pronounced than the bulklike L-edge (squares, $l_0=0$).
  \label{fig_specavg}
  }
\end{figure}

Mesoscopic fluctuations occur in both the photoabsorption $A(\omega)$ and the
excess energy $\omega-\omega_{\rm th}$; the former arise from fluctuations in
the dipole matrix elements and all energy levels, the latter from the levels
above the Fermi energy.  Here we limit ourselves to studying the average values
of $A(\omega)$; fluctuations will be discussed elsewhere \cite{inprep}. We give
$\la A(\omega) \ra$ at the average excitation energy for a given final state,
and measure all energies in mean level spacings $\Deltaml$ from the threshold
$\omega_{\rm th}$ at which the core electron is excited to the perturbed level
$\lam_{M}$ just above the Fermi energy.  The normalized results are in
Fig.~\ref{fig_specavg}(a).  The potential strength $\vc=-10 \Deltaml$ is chosen
so that the Friedel sum rule is fulfilled ($\delta_0 \approx -\pi/2$, $
\delta_{l>0}=0$).  Shown are the spectra for the two spin types
and the total result. We find a peaked edge for the
optically active spin sector as well as for the full spin result (obtained after
convolution with the spectator spin) \cite{remark_singlespec}.

It is interesting to compare the chaotic case to the bulklike situation
(equidistant $\{ \eps \}$ and constant dipole matrix elements). Here, however,
since each orbital channel acts ``independently'', one needs to specify whether
the optically active channel is the one affected by the perturbation. As
mentioned before, we have assumed $s$-type local symmetries, corresponding to a
K-edge ($l_0=1$, $\delta_{l>0}=0$, $\delta_{l=0} \approx -\pi/2$).  Since then only
AOC plays a role, the bulklike situation gives a \textit{rounded} edge; in
contrast, a \textit{peaked} edge is found in the mesoscopic (chaotic) case
[Fig.~\ref{fig_specavg}(b)].  This striking difference is an effect of the
coherent confinement in the chaotic system, in which coupling is to the
derivative of the wave function and so independent of the wave function itself.

This peak is less pronounced, however, than that for the bulklike L-edge case
($l_0 = 0$, so the optically excited and perturbed channels coincide). Because
the dipole matrix elements $w_{i}$ fluctuate in sign and magnitude in the
mesoscopic case, the coherence of the replacement processes is less effective.
Another difference between these two cases concerns their $N$-dependence:
whereas the peak sharpens with increasing $N$ for the bulklike case, it
diminishes with $N$ in the chaotic situation.

The basic physics we study -- Fermi edge singularities due to a localized (rank
one) perturbation acting on a finite number of chaotic electrons via Fermi
golden rule (dipole) matrix elements -- is very general and allows a multitude
of experimental realizations \cite{qwells,localizedIV,inprep}. The technical
requirements for a direct implementation of mesoscopic x-ray absorption
considered here using metallic nanoparticles may be met within a few years.
Another realization would be a double quantum dot with a tiny constriction
\cite{levitov}. Feasible with currently standard semiconductor nanotechnology is
an experiment based on a quantum dot array where the role of the core electron
is taken by a localized electron bound to a (suitable) impurity level in the
band gap. The excitation energy is provided by a micrometer laser that allows
energy resolution well below a mean level spacing. Our results predict a
(slightly) peaked K-edge in the average photoabsorption that becomes rounded for
a bulk two-dimensional electron gas.

We thank K.~Matveev for several helpful discussions and I.~Aleiner, P.~Fuoss,
Y.~Gefen, I.~Lerner, E.~Mucciolo, U.~R\"o{\ss}ler, I.~Smolyarenko,
W.~Wegscheider, and D.~Weiss for useful conversations. M.H. thanks the Humboldt
Foundation for support. This work was supported in part by the NSF
(DMR-0103003).



\begin{thebibliography}{99}
\bibitem{remark_singlespec}
Whereas  the  averaged photoabsorption  has a  peaked  edge,
individual spectra may show ``zig-zagging'' points.
\end{thebibliography}
\end{document}